\documentclass[aps,prb,preprint,superscriptaddress,floatfix,showpacs,amsmath,amssymb]{revtex4-1}

\usepackage[dvips]{graphicx}

\usepackage[colorlinks=true,linkcolor=blue,citecolor=blue]{hyperref}% add hypertext capabilities

\begin{document}

%\preprint{}

\title{Experimental verification of contact-size estimates in point-contact spectroscopy on superconductor/ferromagnet heterocontacts}


\author{J. Gramich}
\affiliation{DFG-Center for Functional Nanostructures, Karlsruher Institut f\"ur Technologie, 76131 Karlsruhe, Germany}
\author{P. Brenner}
\affiliation{DFG-Center for Functional Nanostructures, Karlsruher Institut f\"ur Technologie, 76131 Karlsruhe, Germany}
\author{C. S\"urgers}
\affiliation{Physikalisches Institut, Karlsruher Institut f\"ur Technologie, 76131 Karlsruhe, Germany}
\author{H. v. L\"ohneysen}
\affiliation{DFG-Center for Functional Nanostructures, Karlsruher Institut f\"ur Technologie, 76131 Karlsruhe, Germany}
\affiliation{Physikalisches Institut, Karlsruher Institut f\"ur Technologie, 76131 Karlsruhe, Germany}
\affiliation{Institut f\"ur Festk\"orperphysik, Karlsruher Institut f\"ur Technologie, 76021 Karlsruhe, Germany}
\author{G. Goll}
\email{gernot.goll@kit.edu}
\affiliation{DFG-Center for Functional Nanostructures, Karlsruher Institut f\"ur Technologie, 76131 Karlsruhe, Germany}
%\author{}
%\thanks{}
%\affiliation{}

\date{\today}

\begin{abstract}
Nanostructured superconductor/ferromagnet heterocontacts are studied in the different transport regimes of point-contact spectroscopy. Direct measurements of the nanocontact size by scanning electron microscopy allow a comparison with theoretical models for contact-size estimates of heterocontacts. Our experimental data give evidence that size estimates yield reasonable values for the point-contact diameter $d$ as long as the samples are carefully characterized with respect to the local electronic parameters.
\end{abstract}

\pacs{73.63.Rt, 74.78.Na, 81.07.Lk, 73.23.-b}

\maketitle

\section{Introduction \label{sec:Introduction}}

Point-contact spectroscopy (PCS) has long been known as a method to study the interactions of electrons with other excitations in metals.~\cite{Yanson:1974,Jansen:1980} The interpretation of the observed characteristics in point-contact (PC) spectra is usually difficult because most often contacts are made by the needle-anvil or shear technique and are not  microscopically well-defined with respect to contact size and geometry, structure, and local electronic parameters. Recently, Andreev reflection at point contacts was used to extract values of the transport spin polarization $P$ out of spectra measured on superconductor/ferromagnet (S/F) contacts.~\cite{Soulen:1998,Upadhyay:1998,Perez:2004} However, different models~\cite{Soulen:1998,Upadhyay:1998,Strijkers:2001,Mazin:2001,Perez:2004,Stokmaier:2008} used to describe the transport through S/F interfaces yielded varying values for $P$, also depending on the contact fabrication and the transport regime,~\cite{Woods:2004} an issue that is not yet understood in detail.~\cite{Chalsani:2007} Therefore, a key issue in PCS is to determine the PC parameters, such as the form and diameter of the metallic nanobridge and the mean free path in the immediate contact region, so that one is able to identify the relevant transport regime. Usually Sharvin's \cite{Sharvin:1965} or Wexler's \cite{Wexler:1966} formulae for the ballistic and diffusive transport regime, respectively, are used to infer the PC diameter from the measured PC resistance. Only very few experimental studies deal with the question whether these formulae - especially the interpolation formula in the diffusive regime - yield correct values for the PC diameter. 

In this paper, we employ e-beam lithography to structure a nanometer-sized orifice into a free-standing insulating Si$_{3+x}$N$_{4-x}$ membrane followed by metallization of both sides of the membrane to get a Pb/Fe contact with well-defined orifice size. A detailed characterization of heterocontacts with respect to contact size and geometry, structure, and local electronic parameters allows a direct comparison of the measured PC parameters to different contact-size estimates. We find that the theoretical approximation of the contact size is appropriate if the measured electronic mean free path of each individual contact region is used. The current assignment of the contact regime is facilitated by the analysis of the PC spectrum with features due to electron-phonon interaction or the pair-breaking critical current through the orifice.

\section{Contact models \label{sec:ContactModels}}

When considering electron transport through a circular microscopic constriction with diameter $d$ between two equal metallic reservoirs, different transport regimes have to be distinguished. In the ballistic regime where $d$ is much smaller than the elastic and inelastic electron mean free paths $l_\mathrm{el}$ and $l_\mathrm{in}$ ($d \ll l_\mathrm{in}, l_\mathrm{el}$), the electrons pass the constriction mostly without scattering. The resistance of a PC can be calculated according to Sharvin~\cite{Sharvin:1965} as
\begin{equation}
\label{eq:SharvinRes}
R_\mathrm{Sh}=\dfrac{16\rho l}{3\pi d^2},
\end{equation}
where $\rho l=mv_\mathrm{F}/ne^2$ is a material constant with Fermi momentum $mv_\mathrm{F}$, elementary charge $e$, electron density $n$, and the total mean free path $l$ obtained using Matthiessen's rule. In the opposite case $d \gg l_\mathrm{in}, l_\mathrm{el}$ known as thermal limit where elastic and inelastic scattering takes place in the immediate contact region, the resistance can be calculated after Maxwell~\cite{Maxwell:1904} as
\begin{equation}
\label{eq:MaxwellRes}
 R_\mathrm{M}=\frac{\rho}{d}\,.
\end{equation}
When the transport is mainly diffusive with $l_\mathrm{el} < d \ll \sqrt{l_\mathrm{in}l_\mathrm{el}}$, Wexler~\cite{Wexler:1966} derived an interpolation formula between the two regimes for the contact resistance
\begin{equation}
\label{eq:WexlerRes}
 R_\mathrm{W}=\dfrac{16\rho l}{3\pi d^2}+\gamma\,\dfrac{\rho}{d}\,,
\end{equation}
where the Maxwell term (\ref{eq:MaxwellRes}) is weighted with a slowly varying, non-analytical function $\gamma$. This function can be approximated by using the Pad\'e fit.~\cite{Nikolic:1999}

Based on these formulae, an estimate of the PC diameter $d$ can be obtained as long as the transport regime, the resistance $R$ and the PC parameters of the individual contact  such as the local $l$, and the local resistivity $\rho$ in the contact region are known. Usually these parameters are not determined experimentally but the bulk values found in literature are used instead. For geometrically symmetric heterocontacts of two different metals 1 and 2, $\rho$ and  $\rho l$ are replaced by $(\rho_1+\rho_2)/2$ and by $((\rho l)_1+(\rho l)_2)/2$, respectively.~\cite{NaidyukPCS:2005}

An alternative method to determine the local PC parameters and transport regimes \textit{individually} for \textit{each} PC arises from differentiation of Eq.(\ref{eq:WexlerRes}). For a diffusive PC and under the assumption of dominant phonon scattering one gets
\begin{equation}
  d=\dfrac{\partial\rho_\mathrm{Ph}(T)/\partial T}{\partial R_\mathrm{N}/\partial T}\,,
  \label{eq:Akimenko1}
 \end{equation}
valid in a region where both the PC resistance in the normalconducting state $R_\mathrm{N}(T)$ and the phonon contribution to the resistivity $\rho_\mathrm{Ph}(T)$  have the same functional temperature dependence. Under further assumption that Wexler's formula (\ref{eq:WexlerRes}) is valid and $\gamma\approx 1$, we gain an estimate for the -- at low temperatures dominant -- elastic scattering length in the \textit{immediate} contact region and therefore an independent estimation of the spectroscopic regime:
\begin{equation}
 l_\mathrm{el}\approx\frac{\rho l}{d}\left[R_\mathrm{N}(T=0)-\frac{16\rho l}{3 \pi d^2}\right]^{-1}.
 \label{eq:Akimenko2}
\end{equation}
This approach should yield more reliable values for $d$ and $l_\mathrm{el}$ than Eq. (\ref{eq:WexlerRes}) where the intrinsic resistivity can differ from that determined on reference samples.~\cite{NaidyukPCS:2005} The method was first experimentally verified by Akimenko \textit{et al.}~\cite{Akimenko:1982} for Cu-Cu-homocontacts and the values for $d$ and $l_\mathrm{el}$ obtained from Eqs. (\ref{eq:Akimenko1}) and (\ref{eq:Akimenko2}) could reproduce the intensity of theoretical phonon spectra. 

\section{Experiment \label{sec:Experiment}}
\subsection{Fabrication and characterization of point contacts}\label{FabricationPC}

\begin{figure*}[ht]
	\begin{center}
	\includegraphics[width=0.93\textwidth]{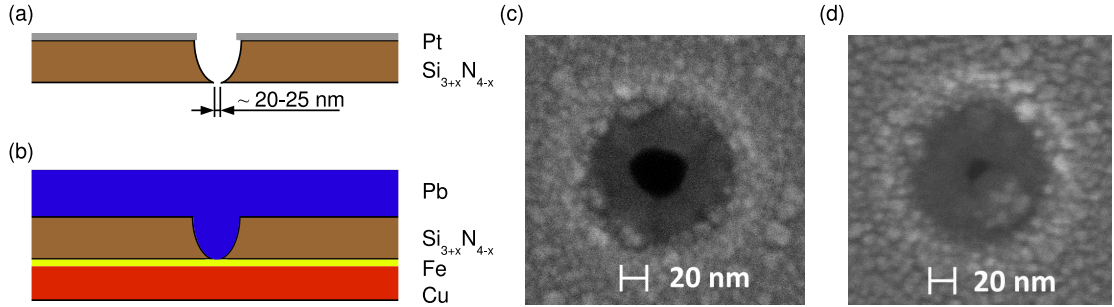}	
\end{center}
	\caption[]{(Color online) (a) Schematic representation of the hole in the silicon nitride membrane. (b) Schematic representation of a Pb/Fe point contact after Pt removal and metallization. (c, d) SEM pictures of etched nanoholes in the membrane with $d\sim 38\,$nm respectively $16\,$nm.}
	\label{fig:Nanoholes}
\end{figure*}

We fabricate PCs by structuring nanobridges between the two metallic reservoirs by means of e-beam lithography. This technique -- originally developed by Ralls~\cite{Ralls:1989} and used in a number of publications~\cite{Upadhyay:1998, Perez:2004, Stokmaier:2008} -- offers a variety of advantages compared to standard PC techniques like the needle-anvil or the shear method. First, mechanically stable contacts with small contact diameters down to a few nanometers can be obtained. Furthermore, \textit{in situ} preparation of the two metal reservoirs yields clean metal interfaces without oxide barriers. The main advantage is, however, that the geometry and size of the contacts are well defined through the orifice size which can be measured by scanning electron microscopy (SEM).

In brief, we follow the process described in Ref.~\onlinecite{Ralls:1989} to fabricate Pb/Fe point contacts, including an additional experimental characterization step. A bowl-shaped hole is structured into a $50-$nm thick insulating non-stoichiometric silicon-nitride membrane (Si$_{3+x}$N$_{4-x}$) by using electron-beam lithography followed by isotropic reactive ion etching with SF$_6$. This leads to a smaller orifice in the membrane than originally structured into the PMMA resist mask. 

\begin{figure}
	\begin{center}
	\includegraphics[width=\columnwidth]{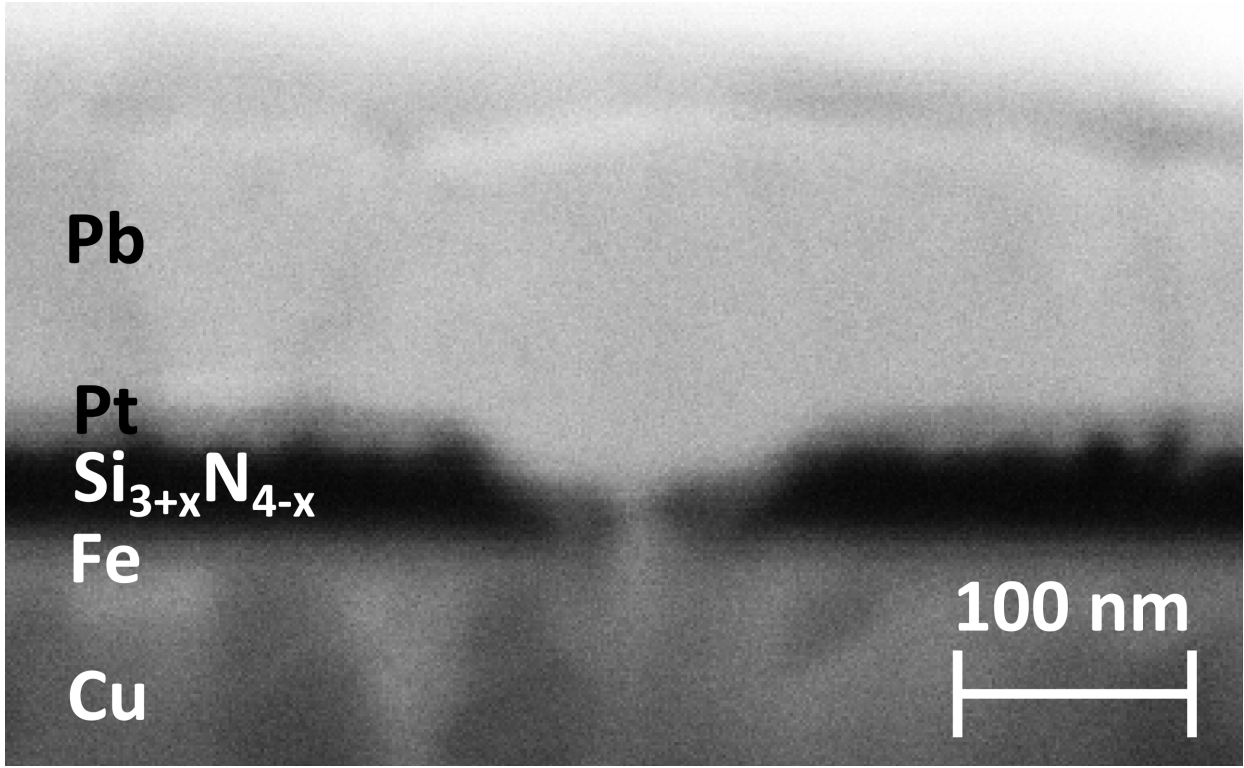}	
\end{center}
	\caption[]{FIB lateral cut through a Pb/Fe point contact of a nominal $24-$nm hole in the membrane. Recorded under $54^\circ$ tilt with SE2 detector and $3\,$kV acceleration voltage.}
	\label{fig:FIBCut}
\end{figure}

The nanoholes in the membranes are analyzed prior to the Pb/Fe metallization by scanning electron microscopy. An approximately $10-20\,$nm thick Pt layer is sputtered onto the membranes. This step is inevitable because charging effects would otherwise destroy the insulating, free-standing membrane and because of the necessary high resolution. We measured the hole diameters in the membrane directly by using a Zeiss Supra 55 VP SEM at small acceleration voltages of $5\,$kV, a $10-\mathrm{\mu}$m aperture and the secondary electron in-lens detector for high surface resolution and contrast.  The silicon-nitride membranes are typically pierced for diameters less than $20\,$nm, with the smallest orifices having a diameter of $\sim 10\,$nm. Figs.~\ref{fig:Nanoholes}(c) and (d) show exemplarily two nanoholes with diameters of $38$ and $16\,$nm. In these SEM pictures, we can clearly distinguish a border between Pt covered areas showing a granular structure and the interior part of the `bowl', which we assume not to be Pt covered. However, further investigations including a focused-ion-beam (FIB) lateral cut through a smaller PC revealed that for smaller nanoholes ($d\lesssim 20-30\,$nm) Pt might reduce the original hole diameter in the membrane by forming a ring of adsorbed material close to the rim of the hole.  Fig.~\ref{fig:FIBCut} shows a FIB lateral cut through a PC with a nominal diameter in the membrane of 24 nm, performed by successive $5\,$nm-distance lateral cuts with a Ga-ion FIB system. One can clearly see the layer geometry with the continuous Pb layer, the "bowl" in the membrane and the corresponding Fe/Cu layer on the back side of the membrane. In addition, one can identify the possible effect of Pt in narrowing the original constriction of the Si$_{3+x}$N$_{4-x}$ membrane. Thus the Pt film is removed completely prior to metallization by immersing the sample in a $T\sim 120\,^\circ$C hot bath of {\it aqua regia} for $\sim 1.5\,$h. Possible organic residues are removed by using a commercial Diener O$_2$ plasma cleaner. According to literature,~\cite{Williams:2003} neither the silicon substrate nor the silicon-nitride membrane are etched by {\it aqua regia} so that the size determination of the nanoholes remains valid. 

In a final step, a $200-$nm thick Pb layer is deposited on the bowl-shaped side of the membrane by e-beam evaporation at room temperature under ultra-high vacuum ($\sim 10^{-9}\,$mbar), followed by a $180^\circ$ {\it in-situ} rotation of the sample and the deposition of a $12-$nm thick Fe layer, topped by a $188-$nm thick Cu layer for a good ohmic contact on the flat side of the membrane (see Fig.~\ref{fig:Nanoholes}(b)). Due to the evaporation at room temperature, the high mobility of the Pb atoms leads to a Stranski-Krastanov-like island-growth mode. Further SEM analysis gave clear evidence for a continous Pb film with average island sizes of $400-500\,$nm, i.\,e. much larger than the typical PC size. Each sample was characterized by measuring its resistance at room temperature in order to check whether a conductive nanobridge has been realized in the metallization process. It turned out that a nanobridge was only established in samples with hole diameters in the membrane larger than $20\,$nm reliably. 9 point contacts with hole diameters ranging from $24-70\,$nm and resistances $R_\mathrm{N}$ in the range of $1.5$ to $34\,\Omega$ were investigated (see also Table~\ref{table:contactradius}).

\subsection{Reference samples}\label{ReferenceSamples}

\begin{figure*}[ht]
	\begin{center}
	\includegraphics[width=\columnwidth]{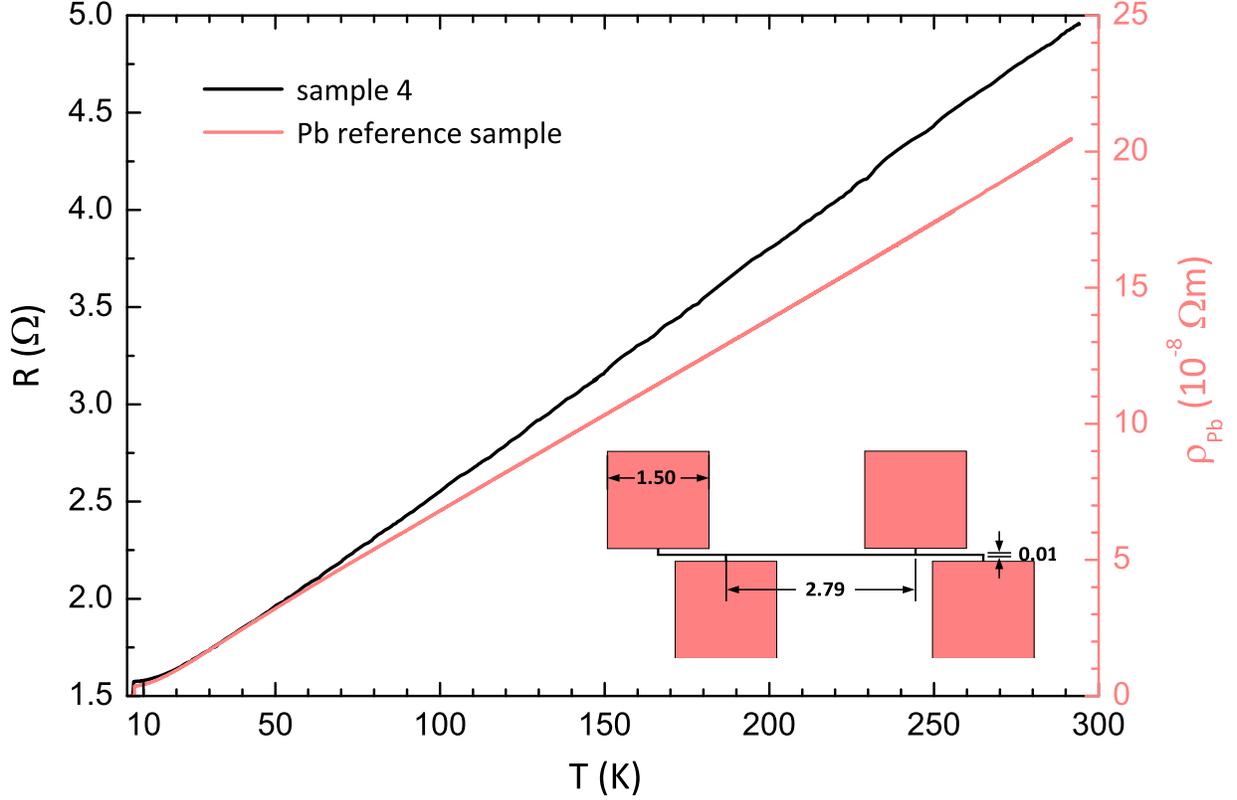}	
\end{center}
	\caption[]{(Color online) Temperature dependence of resistance of the point-contact sample 4 and of the resistivity of the Pb reference sample, respectively. The inset shows a sketch of the reference sample.}
	\label{fig:reference}
\end{figure*}

For an independent determination of the resistivity of the two metallic layers that eventually form the metallic nanobridge we structured reference samples consisting of Pb and Fe layers, respectively, in a usual 4-point measurement geometry by using optical lithography methods and a standard lift-off process. Since the resistivity of thin metallic layers strongly depends on the layer thickness and the growth conditions, the same silicon nitride substrate, identical evaporation conditions and layer thicknesses as for the PCs were chosen. The reference samples had the form of meander-like bars with the dimensions of $\sim 2.8\,$mm length, $\sim 11\,\mathrm{\mu m}$ width (see inset Fig.\,\ref{fig:reference}), and $12\,$nm or $200\,$nm thickness for the Fe and Pb layer, respectively. In addition, the Fe layer was covered {\it in-situ} with a $5-$nm thick insulating SiO$_2$ layer to avoid surface oxidation in air. Resistance measurements were performed instantly after fabrication to reduce oxidation. Sample geometries were determined afterwards by using SEM for the width and length measurements and an Ambios Technology XP-2 profilometer in combination with layer-thickness monitors. 

The resistance $R$ of all samples was measured in the range of room temperature down to $T=1.6\,$K in a $^4$He cryostat by using a LR-700 resistance bridge. The differential resistance $R_\mathrm{d}=\mathrm{d}V/\mathrm{d}I$ vs applied bias voltage $V$ was recorded  at low temperatures between $T\approx 1.5\,$K and $7.3\,$K via lock-in technique. Some measurements were done in an external magnetic field of $\mu_0 H \sim 200 \,$mT applied parallel to the current through the PC to drive the superconducting Pb to the normalconducting state.

\section{Results and Interpretation \label{sec:Results}}
\subsection{Determination of the contact parameters}\label{AnalysisData}

From the $R(T)$ measurements down to $1.6\,$K on the reference samples and PCs (shown in Fig.\,\ref{fig:reference} exemplarily for the Pb reference sample and the PC sample 4) one obtains all necessary contact parameters. For the Pb reference sample we obtain a residual resistance ratio $\mathrm{RRR}=R(300\,\mathrm{K})/R(7.25\,\mathrm{K})\sim 57$, a critical temperature $T_\mathrm{c}\approx 7.24\,$K and the resistance in the normal-conducting state $R_\mathrm{N}=4.020\,\Omega$ at $T=7.25\,$K. With the known geometry of the sample, we calculate the resistivity of our $200-$nm thick Pb layer at $T=7.25\,$K as $\rho^\mathrm{N}_\mathrm{Pb}=R_\mathrm{N}\cdot A/l=(3.596\pm 0.180)\cdot 10^{-9}\,\Omega$m. For our $12-$nm thick Fe layer we obtain $\mathrm{RRR}\sim 1.37$, the averaged resistance $R_\mathrm{N}=7212\,\Omega$ and hence $\rho_\mathrm{Fe}^\mathrm{N}=(3.625\pm 0.363)\cdot 10^{-7}\,\Omega$m at $T=7.25\,$K. Both values agree well with those obtained on similar samples in the literature.~\cite{Zink:2005,Stankiewicz:2006} According to Ref.~\onlinecite{NaidyukPCS:2005}, the resistivity of a geometrical symmetric Pb/Fe heterocontact (HC) at $T=7.25\,$K is
\begin{equation}
 \rho_\mathrm{HC}^\mathrm{N}=\dfrac{\rho_\mathrm{Pb}^\mathrm{N}+\rho_\mathrm{Fe}^\mathrm{N}}{2}=(1.830 \pm 0.182)\cdot 10^{-7}\,\Omega\mathrm{m} \,.
 \label{eq:rhoHC}
\end{equation}
Of course, $\rho_\mathrm{HC}^\mathrm{N}$ is chiefly determined by the highly resistive Fe layer. For further analysis we also need the material constants $\rho l$ for the corresponding metals. Here, we use the value $(\rho l)_\mathrm{Pb}=1.670\cdot 10^{-15}\,\Omega\mathrm{m}^2$ for Pb.~\cite{Aleksandrov:1963} From these values we obtain an electron mean free path in the Pb layer of $l_\mathrm{Pb}=(\rho l)_\mathrm{Pb}/\rho_\mathrm{Pb}^\mathrm{N} \approx 464\,\mathrm{nm}$ at $T=7.25\,$K, which is of the order of the island structure of the Pb film. For Fe, $(\rho l)_\mathrm{Fe}=mv_\mathrm{F}/ne^2=2.652\cdot 10^{-15}\,\Omega\,\mathrm{m}^2$ is calculated using the Drude model with the charge carrier density~\cite{Majumdar:1973} $n=2.65\cdot 10^{28}\,\mathrm{m}^{-3}$ obtained from band-structure and de-Haas-van-Alphen measurements, and the Fermi velocity~\cite{Ashcroft:2005} $v_\mathrm{F}=1.98\cdot 10^6 \,$m/s. The magnitude of the electron mean free path $l_\mathrm{Fe}=(\rho l)_\mathrm{Fe}/\rho_\mathrm{Fe}^\mathrm{N} \approx 7.3\,$nm clearly demonstrates that interface scattering plays the most important role in the $12-$nm thick Fe layer. With these values, we calculate the arithmetic average of $\rho l$ for a heterocontact as
\begin{equation}
 (\rho l)_\mathrm{HC}= \dfrac{(\rho l)_\mathrm{Pb}+(\rho l)_\mathrm{Fe}}{2}=2.161\cdot 10^{-15}\,\Omega\mathrm{m}^2\,.
 \label{eq:rholHC}
\end{equation}
Hence, the electron mean free path through the contact is estimated as $l_\mathrm{HC}=(\rho l)_\mathrm{HC}/\rho_\mathrm{HC}^\mathrm{N} = 11.8\,$nm. These values are usually used to allocate the transport regime.

To calculate PC diameters as suggested by Akimenko \textit{et al.}~\cite{Akimenko:1982} (cf. Eq.\,(\ref{eq:Akimenko1})) one needs additionally $\mathrm{d}R_\mathrm{N}/\mathrm{d}T$ and $\mathrm{d}\rho_\mathrm{HC}/\mathrm{d}T$ of each individual PC as a function of $T$. $\mathrm{d}\rho_\mathrm{HC}/\mathrm{d}T$ is determined by a linear fit for $T>100\,$K to the resistance data of the two reference samples whence 
$\mathrm{d}\rho_\mathrm{HC}/\mathrm{d}T=(\mathrm{d}\rho_\mathrm{Pb}/\mathrm{d}T+ \mathrm{d}\rho_\mathrm{Fe}/\mathrm{d}T)/2=(7.070+6.048)/2\cdot 10^{-10}\,\Omega\,\mathrm{m}/\mathrm{K}=6.559\cdot 10^{-10}\,\Omega\,\mathrm{m}/\mathrm{K}$ in the range where $\rho(T)\sim T$. Similarly, $\mathrm{d}R_\mathrm{N}/\mathrm{d}T$ was determined individually for each PC for $T>100\,$K where $R_\mathrm{N}\sim T$ as well.

\subsection{Comparison of the two different calculation methods with experimental data}\label{ComparisonModels}

Using equations (\ref{eq:SharvinRes}) to (\ref{eq:WexlerRes}) and the parameters given in section~\ref{AnalysisData}, we can determine the contact radius $a=d/2$ and allocate a transport regime to each individual contact. However, this method is problematic because it is based on the assumption of one universal electron mean free path $l_\mathrm{HC}=11.8\,$nm fixed for all PCs. For example, the calculation would result in much larger contact radii $a=55$\,nm respectively 58\,nm for samples 3 and 4, respectively, than the measured upper limits $a_m=33.5$\,nm and accordingly 35\,nm.  In reality, the electron mean free path in the immediate region of the nanobridge may differ from sample to sample due to the granular structure of the films and the growth process of the nanobridge which usually is not controlled on the atomic level. The advantage of the second approach following Akimenko \textit{et al.}~\cite{Akimenko:1982} is that the T dependence of the contact resistance $R_\mathrm{N}$ enters the calculation. It provides a more realistic insight into the dominant scattering processes that determine the PC resistance. We therefore employ Akimenko's~\cite{Akimenko:1982} approach using Eq.~(\ref{eq:Akimenko1}) -- which can be applied nearly independently for contacts in the diffusive regime -- and Eq.~(\ref{eq:Akimenko2}) to calculate the PC parameters $a$ and $l_\mathrm{el}$ for each individual PC with the parameters given in subsection~\ref{AnalysisData}. Table~\ref{table:contactradius} summarizes those values and the allocated transport regimes for all measured samples together with $R_\mathrm{N}$ and the measured hole radii $a_\mathrm{m}$. In addition, the last column displays the PC radii $a_\mathrm{sc}$ calculated by equations (\ref{eq:SharvinRes})-(\ref{eq:WexlerRes}) when taking the correct contact regime and elastic mean free path $l_\mathrm{el}$ determined by Akimenko's approach into account which furnishes a self-consistency check of the method. Indeed, the elastic mean free path in the immediate contact region varies from sample to sample and for some samples differs significantly from $l_\mathrm{HC}$ calculated in the former paragraph. The main reason for the variation is probably a difference in grain structure of the immediate contact region caused by the island growth  of the Pb film which tends to increase the mean free path in larger contacts. The analysis also shows that the low-resistive Pb part of the heterocontact is decisive for the PC properties and prevails over the highly resistive Fe part.

Most of our samples seem to belong to the diffusive (`d') regime with some being closer to the ballistic limit -- labeled as `b' in Table~\ref{table:contactradius} - some being closer to the thermal limit - indicated as `t'. For very small $a$ Wexler's equation (\ref{eq:WexlerRes}) is no longer valid and Eq.~(\ref{eq:Akimenko2}) may yield unphysical negative values of $l_\mathrm{el}$. The corresponding samples can be assigned to the pure ballistic regime (samples 6 and 7). For these samples, the Sharvin formula does indeed yield reliable values for $a$.

\begin{table*}[t]
\begin{center}
{\footnotesize
\begin{tabular}{|c|c|c|c c c|c c|}
\hline 
\multicolumn{2}{|c|}{comparison} & SEM measurement
 & \multicolumn{3}{c|}{calculation with individual mean free path} & \multicolumn{2}{c|}{consistency check}\\ 
\cline{1-2} 
sample no. 
 & $R_\mathrm{N}$ $(\Omega)$
 & $a_\mathrm{m}$ $(\mathrm{nm})$
 & $a$ $(\mathrm{nm})$
 & $l_\mathrm{el}$ $(\mathrm{nm})$
 & transport regime
 & $a_\mathrm{sc}$ $(\mathrm{nm})$
 & used equation
 \\ \hline
 \hline
1
 & $19.60$
 & $17.0$
 & $7.2$
 & $70.4$
 & b
 & $6.8$
 & Sharvin
 \\ 
2
 & $5.48$
 & $31.5$
 & $18.8$
 & $20.1$
 & d
 & $21.1$
 & Wexler
 \\ 

3
 & $1.671$
 & $33.5$
 & $25.3$
 & $178.8$
 & b 
 & $23.4$
 & Sharvin
 \\ 
4
 & $1.575$
 & $35.0$
 & $26.2$
 & $171.3$
 & b
 & $24.1$
 & Sharvin
 \\ 
5
 & $33.48$
 & $12.0$
 & $5.6$
 & $40.6$
 & b
 & $5.2$
 & Sharvin
 \\ 
6
 & $24.95$
 & $20.0$
 & $5.5$
 & $-$
 & b
 & $6.1$
 & Sharvin
 \\ 
7
 & $15.67$
 & $12.0$
 & $6.7$
 & $-$
 & b
 & $7.7$
 & Sharvin
 \\ 
8
 & $12.62$
 & $19.5$
 & $10.7$
 & $22.1$
 & d
 & $11.6$
 & Wexler
 \\ 
9
 & $3.830$
 & $34.5$
 & $23.2$
 & $21.9$
 & t
 & $23.9$
 & Maxwell
 \\ 
\hline 
\end{tabular}}
\end{center}
\caption[]{Calculated point-contact radii $a$ according to Akimenko's approach in comparison to the orifice radii $a_\mathrm{m}$ in the membranes determined by scanning electron microscopy. For the allocation of the transport regimes (b=ballistic, d=diffusive, t=thermal) we used the individual local elastic mean free path $l_\mathrm{el}$ of the point contacts determined by Eq.~(\ref{eq:Akimenko2}). For the calculation of $a_\mathrm{sc}$ in the consistency check we used the indicated resistance formulae with the individual $l_\mathrm{el}$ and the parameters given in subsection~\ref{AnalysisData}.}
\label{table:contactradius}
\end{table*}

\begin{figure}[ht]
	\begin{center}
	\includegraphics[width=\columnwidth]{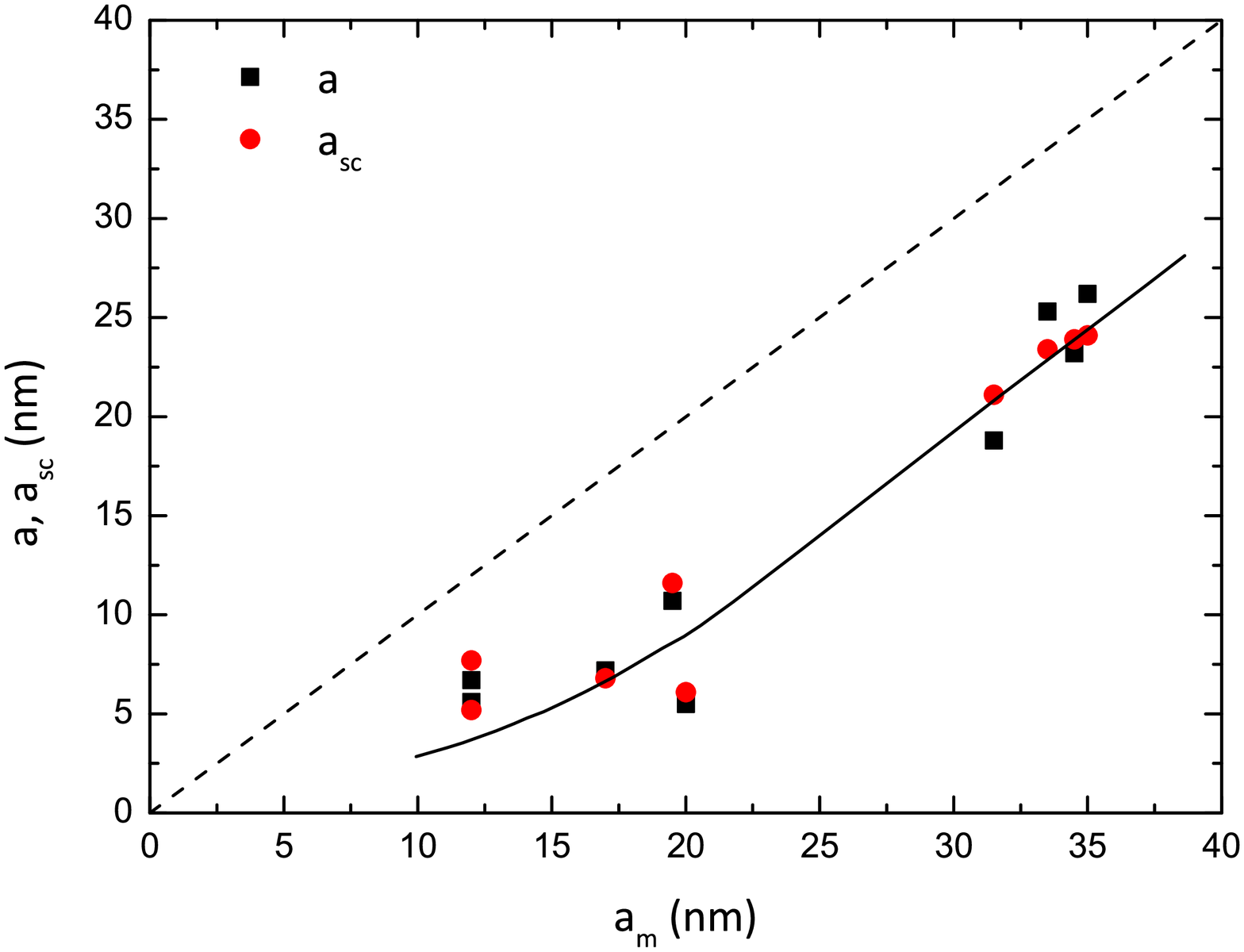}	
\end{center}
	\caption[]{Comparison of the calculated PC radius $a$ according to Akimenko \textit{et al.} \cite{Akimenko:1982} and $a_\mathrm{sc}$ according to the consistency check with the experimentally determined hole radius in the membranes $a_\mathrm{m}$. Dashed line indicates the "ideal" expectation $a=a_\mathrm{sc}=a_\mathrm{m}$. Solid line presents a guide to the eye of experimental data.}
	\label{fig:ComparisonRadius}
\end{figure}

As can be seen from Fig.~\ref{fig:ComparisonRadius}, the contact radii $a_\mathrm{sc}$ derived from the self-consistency check yield nearly the same values for the PC radius $a$ when taking the correct transport regime and the locally determined $l_\mathrm{el}$ into account. The calculated radii are always somewhat smaller than those of the holes in the membranes, but do seem to follow the same general experimental trend. When recalling the sample fabrication process, it seems obvious that the metals will not completely fill the holes in the membranes during the evaporation process - thus leading to narrower \textit{metallic} nanobridges. The SEM measurement yields an upper limit for the effective PC radius. Statistical variations of the contact size can be attributed due to the rather uncontrolled nature of the aggregation process at the atomic level following the evaporation. Hence, the exact geometry of establishing a nanobridge cannot be controlled. 

Finally, we want to emphasize that experimentally determined values only enter into the determination of the PC radius. The very good agreement of those calculations with the experimental trend shows \textit{directly} for metallic heterocontacts that frequently used size estimates for PC diameters do agree with the experimental data and yield reasonable values for those parameters. However, extreme care has to be taken when characterizing the samples.

\subsection{Additional supporting results}\label{SupportResults}

\begin{figure}
	\begin{center}
	\includegraphics[width=\columnwidth]{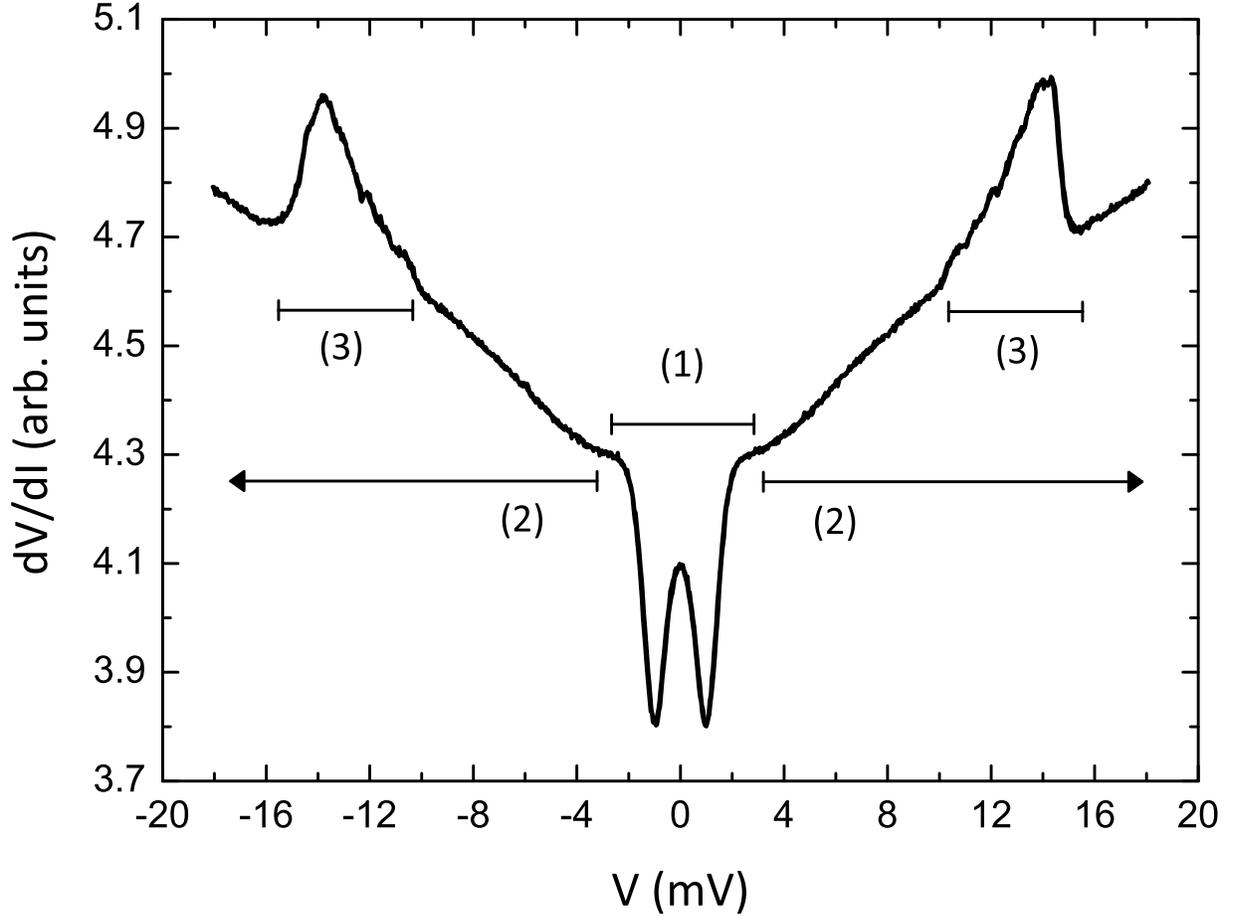}	
\end{center}
	\caption[]{Point-contact spectrum of sample 2 at $T=1.6\,$K. In addition to the Andreev signature (1) in the middle part of the spectrum, we observe a rise and change of slope (2) of $R_d$ as well as a sharp peak (3) on this rise.}
	\label{fig:OverviewSpectr}
\end{figure}

PC spectra have been recorded at different temperatures and in an applied magnetic field $\mu_0 H \sim 200 \,$mT which proved to be sufficient to drive the superconducting Pb into the normalconducting state at lowest $T=1.6\,$K. These spectra reveal clear nonlinearities at voltages $V>\Delta/e$ in the normal and superconducting states of Pb, where $\Delta$ denotes the superconducting energy gap. Fig.~\ref{fig:OverviewSpectr} displays a typical spectrum with Pb in the superconducting state, where we have marked the regimes of nonlinearities. Besides the well-known Andreev double-minimum structure (1) at $|V|=\Delta /e=1.3$\,mV for Pb, we observe an overall rise in the differential resistance and changing slopes which can be interpreted as due to electron-phonon scattering in Pb (2), and a rather sharp peak in the superconducting state of Pb at higher energies $eV\gg 2\Delta$ that scales with temperature is attributed to the current through the contact exceeding the pair-breaking critical current (3). Features of Andreev reflection with the possibility to extract $\Delta$ can be found even for rather ill-defined PCs. The nonlinearities caused by (2) and (3), on the other hand, depend sensitively on the quality of the PC. 

\begin{figure}
	\begin{center}
	\includegraphics[width=\columnwidth]{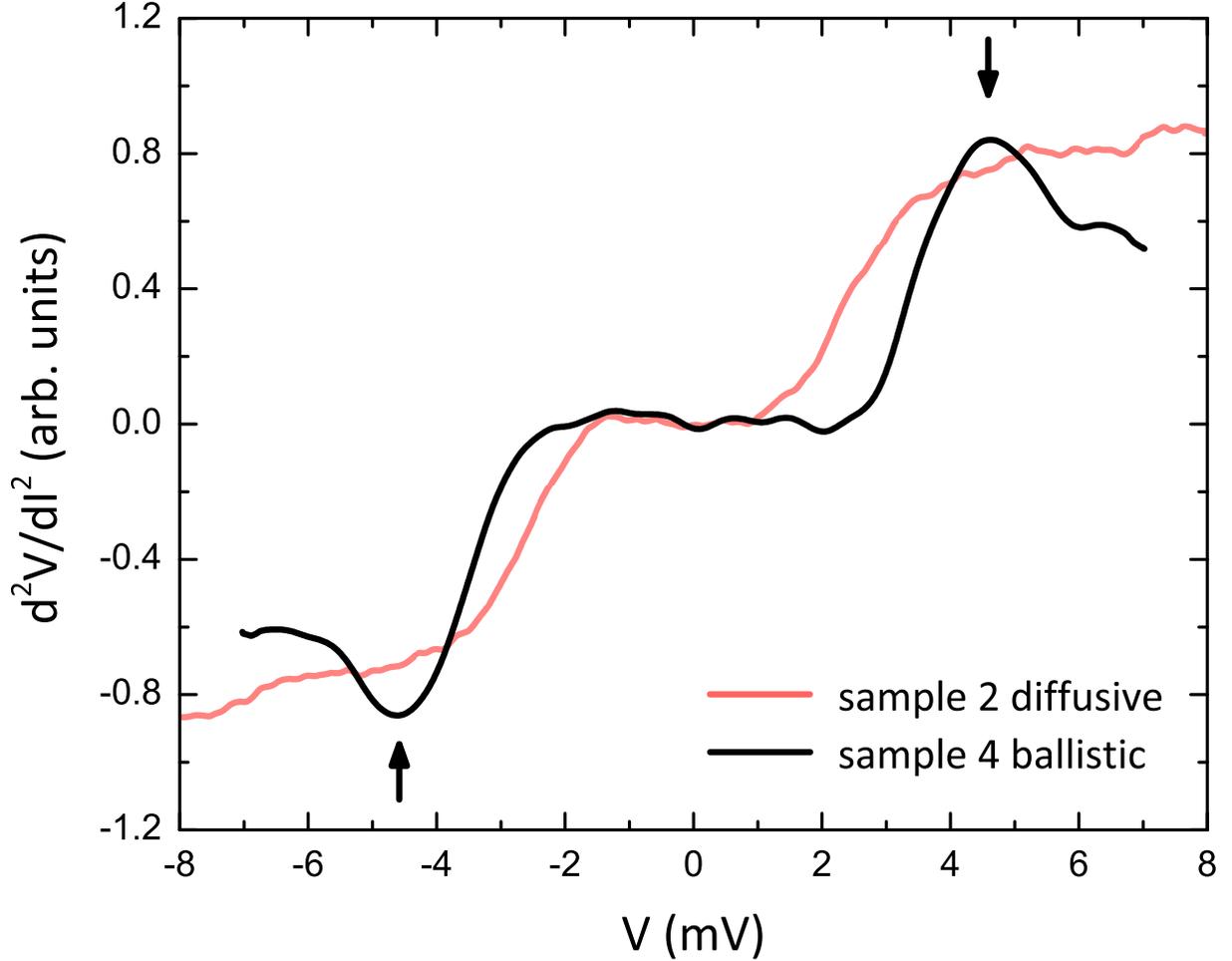}	
\end{center}
	\caption[]{(Color online) Phonon spectra of samples 2 and 4 for $T=1.6\,$K and $\mu_0 H= 200 \,$mT in the normalconducting state, obtained by numerical differentiation $\mathrm{d}R_\mathrm{d}/\mathrm{d}V$. Arrows indicate the position of TA phonon peaks in Pb.}
	\label{fig:Phononspectra}
\end{figure}

Fig.~\ref{fig:Phononspectra} exemplarily shows the second derivative $\mathrm{d}^2V/\mathrm{d}I^2$ vs. $V$ spectra of samples 2 and 4, recorded at $T=1.6\,$K and $\mu_0 H= 200 \,$mT where Pb is in the normal state. $\mathrm{d}^2V/\mathrm{d}I^2$ vs. $V$ spectra are obtained by numerical differentiation of the measured $\mathrm{d}V/\mathrm{d}I$ data. Typically, the data have to be averaged over 20-30 data points to reduce the noise. The second derivative $\mathrm{d}^2V/\mathrm{d}I^2$ is proportional to the Eliashberg function $\alpha_\mathrm{PC}^2F(\omega)$, where for heterocontacts the spectrum is a sum of contributions of both Pb and Fe.~\cite{NaidyukPCS:2005} For low energies, predominantly the phonon excitations of Pb are expected while the contributions of Fe will be significant at higher energies $eV \gtrsim 15-20$\,meV only.~\cite{FootnotePhonons,Lysykh:1980a} Indeed, we identify features indicated by arrows for sample 4 at $|V|\approx 4.5\,$mV, which are ascribed to transverse acoustic (TA) phonons of Pb.~\cite{Yanson:1974} In contrast, only a broad feature is seen in the spectrum of sample 2. The broadening is a direct consequence of the reduced mean free path, corresponding to a Knudsen ratio $K=l_\mathrm{el}/a\approx 1$, of this particular sample. In the diffusive regime, the intensity of the peaks depends on the Knudsen ratio.~\cite{Lysykh:1980b} Indeed, for sample 4 which is closer to the ballistic regime with much larger $K\approx 6.5$ the phonon peak is much more pronounced. 

We note that we observe a shift of the peak positions to higher energies on the order of the energy gap $\Delta\sim 1.3\,$meV and a  peak intensity which is nearly independent of the contact regime when Pb is in the superconducting state, in agreement with theory~\cite{Khlus:1983,Omelyanchouk:1988} and earlier experiments.~\cite{Khotkevich:1990}

\begin{figure}
	\begin{center}
	\includegraphics[width=\columnwidth]{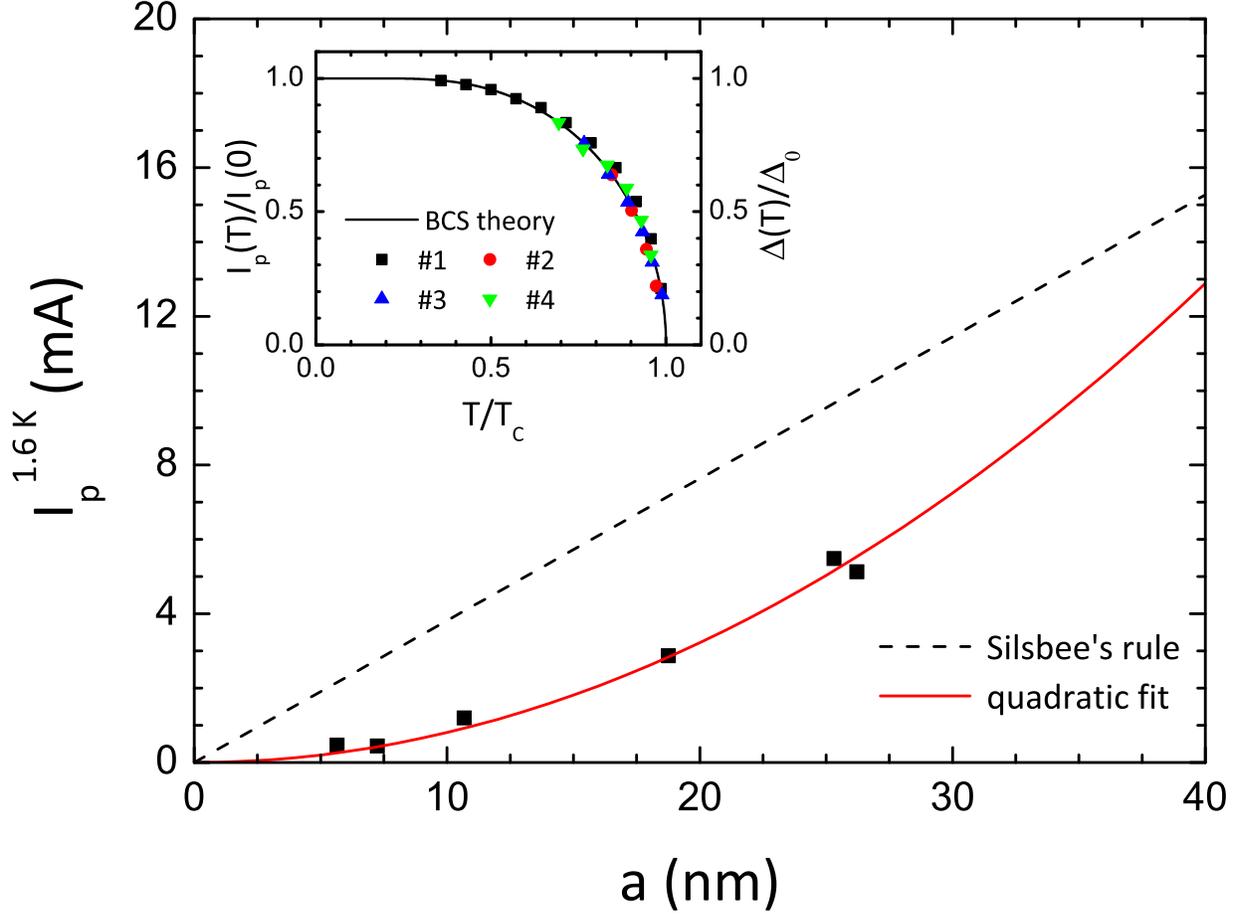}	
\end{center}
	\caption[]{(Color online) Critical current $I_p$ as a function of point-contact radius $a$ at $T=1.6$\,K and temperature (inset). The dashed line indicates the expected linear dependence according to Silsbee's rule, the solid red line is a quadratic least-squares fit of the data of six point contacts in the ballistic or diffusive limit. The inset compares the temperature dependence of the peak current for four samples with the BCS temperature dependence of the order parameter. The experimental data point at lowest temperature of each sample has been scaled on the BCS curve.}
	\label{fig:CritCurrent}
\end{figure}

The second feature which supports the allocation of the transport regimes is the sharp peak in the differential resistance spectra at higher bias $|V|\gg\Delta/e$ when Pb is in the superconducting state (see Fig.~\ref{fig:OverviewSpectr}, (3)). It is observed for all our PCs. The dc current $I_\mathrm{p}(T)$ at the peak position, simultaneously recorded as a function of $T$, scales with the superconducting order parameter in BCS theory (see inset in Fig.~\ref{fig:CritCurrent}, where the $T$ dependence is studied for samples 1-4). Therefore, we identify the peak current $I_p$ with the critical current $I_\mathrm{crit}$. The sharp peaks arise when the current through the contact exceeds the pair-breaking critical current density of Pb, leading to a sudden rise in the differential resistance. 

The appearance of those peaks in the spectra of PCs with a superconducting counterelectrode has been analyzed in a number of publications.~\cite{Naidyuk:1991,Lysykh:1992,Haeussler:1996,Sheet:2004} The peaks have been interpreted, e.\,g., as being akin to PCs with large contact dimensions,~\cite{Sheet:2004} i.\,e. contacts in the thermal regime. However, that this is not necessarily the case. The position $V_\mathrm{p}$ of the peaks varies depending on the contact geometry and can often be found at voltages as high as $V\sim 20$\,mV at lowest $T$. As discussed above, our fabrication process results in a controlled PC geometry, where -- in our specific sample geometry -- Pb establishing the contact grows in a cylindrical shape into the nanohole (see Fig.~\ref{fig:FIBCut}). 

According to Silsbee's rule for a cylindrical superconductor with radius $a \gg \lambda_\mathrm{L}$ ($\lambda_\mathrm{L}$: London penetration depth) superconductivity is destroyed if the self-field at the surface produced by the current through the wire reaches the thermodynamic critical field $B_\mathrm{c}^\mathrm{th}$. It follows that $I_\mathrm{crit}= (2\pi B_\mathrm{c}^\mathrm{th} a)/\mu_0 $, where $\mu_0$ is the permeability of free space, i.\,e. $I_\mathrm{crit}\propto a$ (Ref.\,~\onlinecite{Buckel:2004}). For $a<\lambda_\mathrm{L}$, which is the case for our contacts, one would expect a geometry independent and constant pair-breaking current density $j_\mathrm{crit}$. MacDonald and Leavens~\cite{MacDonald:1983} showed that the current distribution through the contact depends on the ratio $a/l_\mathrm{el}$. While for clean contacts with $a\lesssim l_\mathrm{el}$ the current density is practically constant over the cross-section, it increases abruptly at the periphery for dirty contacts with $a\gg l_\mathrm{el}$. Indeed, several experimental studies~\cite{Lysykh:1992,Haeussler:1996} have shown that the critical current for contacts in the thermal transport regime scales as $I_\mathrm{crit}\propto a$, while for clean contacts $I_\mathrm{crit}\propto a^2$ was found in accordance with Ref.~\onlinecite{MacDonald:1983}. Fig.~\ref{fig:CritCurrent} shows the critical current $I_p$ at $T=1.6\,$K vs. the PC radius $a$ for six of our nearly ballistic or diffusive contacts. The critical current of our contacts does not follow Silsbee's rule shown as a dashed line\cite{FootnoteCritCurrentCalc} but rather shows a quadratic dependence. A least-squares fit to the data via $I_\mathrm{p}(T=1.6\,\mathrm{K})=j_\mathrm{crit}(T=1.6\,\mathrm{K})\cdot \pi a^2$ yields an universal critical current density for all contacts of $ j_\mathrm{crit}(T=1.6\,\mathrm{K})=(2.56\pm 0.09)\cdot 10^8\, \mathrm{A}/\mathrm{cm}^2$. This value is of the same order of magnitude as the calculated~\cite{FootnoteCritCurrentCalc} BCS value for Pb $j_\mathrm{crit}^\mathrm{theo,BCS}(T=0)=(4B_\mathrm{c}^\mathrm{th})/(3\sqrt{6}\mu_0 \lambda_\mathrm{L})=8.9\cdot 10^7 \,\mathrm{A}/\mathrm{cm}^2$ and an experimentally determined value~\cite{Hunt:1966} on $50-\mathrm{nm}$ thick Pb layers of $j_\mathrm{crit}^\mathrm{exp}(T=0\,\mathrm{K})=5.26\cdot 10^7 \,\mathrm{A}/\mathrm{cm}^2$. The observed behavior and the geometry independent critical current density for all contacts with $a<l_\mathrm{el}$ confirms our assumption that the observed peaks arise from reaching the pair-breaking current in the immediate contact region and supports our assignment of the transport regimes and the calculated PC radii.

\section{Conclusion \label{sec:Conclusion}}

In conclusion, we have presented an experimental study of size estimates for heterocontacts in PC spectroscopy in the different transport regimes. A direct SEM measurement of the nanocontact size allows a comparison with theoretical models for contact-size estimates of heterocontacts in the semiclassical approach. Due to the good agreement between experimental and calculated values, we conclude that the semiclassical models yield reasonable values for the PC diameter $d$ as long as the samples are carefully characterized and the correct transport regime is determined taking the local transport parameters of the individual contact into account. Our assignment of the samples to different transport regimes is corroborated by the analysis of further features in the spectra such as phonon peaks and the critical pair-breaking current of Pb.

\begin{acknowledgments}
The authors would like to thank E. M\"uller, S. K\"uhn and T. Peichl for stimulating discussions of the sample fabrication process and C. Reiche for the support in the fabrication of the reference samples. We acknowledge the financial support provided within the DFG-Center for Functional Nanostructures.
\end{acknowledgments}


\begin{thebibliography}{99}
\bibitem{FootnotePhonons}
Due to experimental limitations of our measurement setup (change in the measuring range of our precision current source), some PCs with lower contact resistances could only be measured up to voltages of $6-8\,$mV. However, since phonon excitations for Pb should already be visible at relatively low energies, this data seems to be sufficient for a qualitative discussion. The longitudinal acoustic (LA) phonon which is expected as a peak around $|V|\approx 8$\,mV  is barely visibly for sample 2 but often covered by the additional raise of the critical-current peak.
\end{thebibliography}
\end{document}